\documentclass{class_nature}

\usepackage[utf8]{inputenc}
\usepackage[T1]{fontenc}
\usepackage{textcomp}
\usepackage{gensymb}
\usepackage{amsmath} 
\usepackage{xcolor}
\usepackage{graphicx}
\usepackage{xfrac}

\makeatletter
\DeclareRobustCommand\bfseriesitshape{%
  \not@math@alphabet\itshapebfseries\relax
  \fontseries\bfdefault
  \fontshape\itdefault
  \selectfont
}
\makeatother
\DeclareTextFontCommand{\textbfit}{\bfseriesitshape}

\newcommand{\FefSis}{Fe$_{40}$Si$_{60}$}
\newcommand{\FefSif}{Fe$_{45}$Si$_{55}$}
\newcommand{\FeSi}{Fe$_{x}$Si$_{1-x}$}

\title{Spin-orbit torque generated by amorphous \FeSi}

\author
{Cheng-Hsiang Hsu$^{1\ast}$, Julie Karel$^{2, 3\ast}$, Niklas Roschewsky$^{4}$, Suraj Cheema$^{5}$, Dinah Simone Bouma$^{4,6}$, Shehrin Sayed$^{1}$, Frances Hellman$^{4,5,6}$, Sayeef Salahuddin$^{1,6}$
}

\begin{document} 

\maketitle

\begin{affiliations}
 \item Department of Electrical Engineering and Computer Science, University of California, Berkeley, California 94720, USA
 \item Department of Materials Science and Engineering, Monash University, Clayton, Victoria 3800, Australia
 \item Australian Research Council Centre of Excellence in Future Low-Energy Electronics Technologies, Monash University, Clayton, Victoria 3800, Australia
 \item Department of Physics, University of California, Berkeley, California 94720, USA
 \item Department of Materials Science and Engineering, University of California, Berkeley, California 94720, USA
 \item Materials Science Division, Lawrence Berkeley National Laboratory, Berkeley, California 94720, USA
 \newline \textbf{*These authors contributed equally to this work}
\end{affiliations}

\newpage
%%%%%%%%%%%%%%%%% END OF PREAMBLE %%%%%%%%%%%%%%%%

%%%%%%%%%%%%%%%%%%%%%%%%%%%%%%%%%%%%%%%%
%    Abstract/summary 
%%%%%%%%%%%%%%%%%%%%%%%%%%%%%%%%%%%%%%%%
\begin{abstract}
While tremendous work has gone into spin-orbit torque\cite{Liu2011, Miron2011, Liu2012, Mellnik2014, Hayashi2014, Avci2014, Wang2015, Chen2016, Skinner2015, Shao2016, MacNeill2017, Nan2019, Manchon2019, Luo2020} and spin current generation\cite{Noel2020, Tsai2020}, charge-to-spin conversion efficiency remains weak in silicon\cite{Ando2012} to date, generally stemming from the low spin-orbit coupling (low atomic number, Z) and lack of bulk lattice inversion symmetry breaking. Here we report the observation of spin-orbit torque in an amorphous, non-ferromagnetic \FeSi\ / cobalt bilayer at room temperature, using spin torque ferromagnetic resonance\cite{Liu2011, Liu2012, Mellnik2014, Wang2015, Skinner2015, Chen2016, MacNeill2017, Nan2019} and harmonic Hall measurements\cite{Hayashi2014, Shao2016, Avci2014, Roschewsky2017, Roschewsky2019}. Both techniques provide a minimum spin torque efficiency of about 3 \%, comparable to prototypical heavy metals such as Pt\cite{Liu2011} or Ta\cite{Liu2012}. According to the conventional theory of the spin Hall effect\cite{Hirsch1999,Sinova2015, Liu2012}, a spin current in an amorphous material is not expected to have any substantial contribution from the electronic bandstructure. This, combined with the fact that \FeSi\ does not contain any high-Z element, paves a new avenue for understanding the underlying physics of spin-orbit interaction and opens up a new class of material systems - silicides - that is directly compatible with complementary metal-oxide-semiconductor (CMOS) processes\cite{Zhang2003, Chen2005, Ren2011} for integrated spintronics applications. 
\end{abstract}

\clearpage

%%%%%%%%%%%%%%%%%%%%%%%%%%%%%%%%%%%%%%%%%%%%%%%%%%%%
% Intro to SOT, materials demonstrated to exhibit SOT (weak in Si)
%%%%%%%%%%%%%%%%%%%%%%%%%%%%%%%%%%%%%%%%%%%%%%%%%%%%
Spin-orbit torque (SOT) induced magnetization switching has shown great potential for next-generation magnetic memory technologies\cite{Salahuddin2018}. Intense effort is currently in place to explore new materials for and new mechanisms of SOT as well as its use for next generation logic computing and beyond\cite{Luo2020, Noel2020, Tsai2020}. Conventionally, when an in-plane current is applied through the non-magnetic underlayer with strong spin-orbit interaction in a non-magnetic / magnetic heterostructure, a transverse spin current is generated. There are two main physical mechanisms responsible for the spin current: (i) spin Hall effect (SHE) stemming from Berry curvature of the bulk bandstructure or scattering off heavy (large atomic number, Z) impurities\cite{Liu2011, Dyakonov1971, Hirsch1999, Sinova2015} and (ii) the Rashba-Edelstein effect (REE) due to broken inversion symmetry at the interface\cite{Miron2011, Edelstein1990, Hellman2017}. The spin current in turn exerts a torque on an adjacent magnetic layer \cite{Slonczewski1996, Ralph2008, Manchon2008}. In most material systems, these two physical mechanisms can act in parallel, and it has been challenging to differentiate their separate contributions quantitatively\cite{Manchon2019}.

Spin-orbit torque has been demonstrated in various non-magnetic materials including heavy metals\cite{Liu2011,Miron2011,Liu2012}, topological insulators\cite{Mellnik2014,Wang2015}, semiconductors\cite{Chen2016,Skinner2015}, two dimensional materials\cite{Shao2016, MacNeill2017}, and three dimensional semi-metals\cite{Nan2019}. Beyond these prototypical materials, here we report the observation of spin-orbit torque in an amorphous, non-ferromagnetic iron silicide (a-\FeSi) and cobalt (Co) bi-layer at room temperature. By injecting an in-plane charge current in a-\FeSi, a torque is exerted on the magnetic moments in the adjacent cobalt layer (Fig. 1a). This torque results in an appreciable modulation of the magnetization that is detected electrically via the spin torque ferromagnetic resonance technique and harmonic Hall measurement. Both techniques yield a minimum spin torque efficiency of approximately 3 \%, which is the same order of magnitude as platinum (8 \%)\cite{Liu2011} and 300 times larger than the spin Hall angle (spin torque efficiency) reported in crystalline, p-type silicon (0.01 \%)\cite{Ando2012}. In contrast to the trends observed conventionally, substantial spin orbit torque from an amorphous, low-Z material indicates that a reasonably strong spin current can be generated in a material lacking a bandstructure or heavy element. 

%%%%%%%%%%%%%%%%%%%%%%%%%%%%%%%%%%%%%%%%%%%%%
%\paragraph*{Entering the scope of the study}
%%%%%%%%%%%%%%%%%%%%%%%%%%%%%%%%%%%%%%%%%%%%%

\FeSi, along with many other transition metal silicides, has been utilized extensively in the semiconductor industry. For decades, silicides have been instrumental in forming electrical contacts, gate electrodes, local interconnects and diffusion barriers\cite{Chen2005} in very large scale integration (VLSI) technologies. Moreover, self-aligned silicidation (SALICIDE) remains a critical process in ensuring low contact and series resistance to the source and drain
region of a planar metal-oxide-semiconductor field-effect transistor (MOSFET)\cite{Zhang2003, Ren2011}. Metal
silicides remain a crucial technology as the down scaling of transistors progresses, enabling
better performance and denser packing of devices on an integrated circuit. In our \FeSi\ films, the silicon composition ranges between 55 \% and 60 \%. The amorphous nature of \FeSi\ films with similar concentration has been studied extensively\cite{Sharma1975, Karel2018}, and the amorphous structure in our devices was further confirmed by high-resolution cross-sectional transmission electron microscopy (HR-TEM) and synchrotron grazing incidence X-ray diffraction performed directly on the bi-layers. In the HR-TEM image (Fig. 1b), no nanocrystal formation or evidence of lattice fringes are observed across the \FeSi\ layer. Additional TEM based structural characterizations of \FeSi\ such as local fast Fourier transform also confirm its amorphous phase (Extended Data Fig. 1). In Fig. 1c, the X-ray 2D pole figure shows the diffraction pattern of a 10 nm \FefSif\ film. A highly diffused pattern is observed and only sharp peaks and Kikuchi lines corresponding to the single crystal silicon substrate are present. By contrast, the \FefSif\ / Co bilayer contains additional reflections from the thin crystalline cobalt layer (Fig. 1d). The absence of a diffraction signal from the silicide layer confirms the lack of long-range order or nanocrystals in the \FeSi\ thin films (Extended Data Fig. 2). 

Our bilayers constitute approximately 6 to 10 nm of a-\FeSi\ and 4 nm of cobalt, grown on amorphous silicon nitride on silicon substrates with oxidized aluminum as a capping layer to prevent the cobalt surface from oxidizing (Fig. 1a). a-\FeSi\ and cobalt are grown by electron beam evaporation with controlled composition of iron and silicon via co-evaporation; compositions were verified using Rutherford Backscattering Spectrometry. Aluminum, which is subsequently oxidized by exposure to air, is deposited from an effusion cell. The cobalt magnetic moments are oriented in-plane (Extended Data Fig. 3), and the a-\FeSi\ layer at both compositions (x=40 and 45 \%) show no appreciable ferromagnetism (Extended Data Fig. 4). For device fabrication, we use Ar ion-milling and optical lithography to pattern the film stack into desired device structures.

To characterize the SOT in a-\FeSi\ / Co, we first performed harmonic Hall measurements of the effective magnetic fields due to SOT\cite{Hayashi2014, Avci2014, Shao2016, Roschewsky2017, Roschewsky2019}. The samples are patterned into a single Hall bar structure (Fig. 2b) and an a.c. current is injected along the x-axis (Fig. 2a). Both first ($V_{1\omega}$) and second ($V_{2\omega}$) harmonic transverse voltages are detected (Fig. 2c, d). During the measurement, the equilibrium direction of the Co magnetic moments is pinned by a fixed in-plane external magnetic field as the sample is rotated. We conduct this in-plane angle scan as a function of both current and magnetic field magnitude. If a SOT is present, a quasi-static oscillation of the cobalt moments can be induced and hence an oscillating Hall voltage. In this case, a Hall voltage at twice the injected a.c. current frequency will be excited\cite{Hayashi2014, Avci2014} (Fig. 2d). In addition to the SOT contribution, $V_{2\omega}$ also includes signals induced from the thermoelectric effects due to joule heating of the oscillating current that scales quadratically with current. Thus, thermal effects such as ordinary Nernst effect (ONE) and anomalous Nernst effect (ANE) can be present in $V_{2\omega}$ and extracting the SOT effect requires careful analysis\cite{Roschewsky2019} (see also Methods, Extended Data Fig. 5).

%%%%%%%%%%%%%%%%%%%%%%%%%%%%%%%%%%%%%%%%%%%%%%%%%%%%%%%%%%%%
%%%%%%%%%% Results on harmonic + ST-FMR 
%%%%%%%%%%%%%%%%%%%%%%%%%%%%%%%%%%%%%%%%%%%%%%%%%%%%%%%%%%%%
After doing the comprehensive angle ($\varphi_{B}$), current and magnetic field dependent sweeps, we carefully analyzed $V_{2\omega}$. First, the coefficient of the cos(2$\varphi_{B}$)cos($\varphi_{B}$) component is extracted, giving the field (FL) -like torque effective field voltage that scales as $1/B_{ext}$ (Extended Data Fig. 5b). The FL-like torque effective field includes both the FL-like spin-orbit torque ($\Delta B_{FL, SOT}$) and FL-like torque due to Oersted field ($B_{Oe}$) contribution. By filtering out the Oersted field, strong FL-like SOT effective fields are still present (Fig. 2e). Next, we are left with a cos($\varphi_{B}$) term which contains three contributions including the damping (DL) -like torque effective field ($\Delta B_{DL}$), ANE and ONE (Extended Data Fig. 5a). By considering the magnetic field dependent sweep, we can distinguish between these three contributions separately since the DL-like torque effective field scales as $1/(B_{ext} + \mu_{0} M_{eff}$) (Extended Data Fig. 5c), the ANE is constant with external field and the ONE scales linearly with external field\cite{Roschewsky2019}. In our data, clear signatures of both the DL-like and FL-like effective fields due to spin-orbit torque are present (Fig. 2e, f). Both the field-like and damping-like effective fields scale linearly with the injected current densities in the \FeSi\ layer, confirming the presence of both DL-like and FL-like SOTs (Fig. 2e, f).

We further confirmed and characterized spin-torque strength by spin-torque ferromagnetic resonance (ST-FMR) which, in the past, has been used to quantify SOT strength in heavy metals, topological insulators and semi-metals\cite{Liu2011,Liu2012,Mellnik2014,Wang2015,Skinner2015, Chen2016,MacNeill2017, Nan2019}. The schematic of the measurement setup is shown in Fig. 3a. We fabricated 3 x 9 micron strips and contacted them with co-planar waveguides (Fig. 3b) for radio-frequency (RF) current injection. The injected in-plane RF microwave current induces a torque that acts on the magnetization of the Co layer, yielding an oscillating anisotropic magnetoresistance (AMR). By sweeping an in-plane field, the criteria of ferromagnetic resonance can be fulfilled and a resonance line shape can be obtained by measuring the d.c. mix down voltage ($V_{mix}$) stemming from the mixing between the oscillating AMR and RF current. From the line shape of $V_{mix}$ versus external field, symmetric and anti-symmetric components are extracted which respectively correspond to the in-plane  ($\hat{m}\times(-\hat{y}\times\hat{m})$, $\parallel$) and out-of-plane ($-\hat{y}\times\hat{m}$, $\perp$) component of the current-induced torques (Fig. 3a).

The resulting ($V_{mix}$) line shapes measured at frequencies from 6.5 GHz to 9 GHz shows a strong presence of resonance due to the SOT in the a-\FefSif\ (8) / Co (4) bilayer (Fig. 3c). The magnetic field is oriented in-plane with angle of $\varphi_{B}$ = 45° from the current direction. To analyze the details of various components of the SOT, we fitted both a symmetric and an antisymmetric Lorentzian function (Methods) to the line shape. A typical fit is shown at 6.5 GHz in Fig. 3d. Both symmetric and anti-symmetric components are clearly present.

%%%%%%%%%%%%%%%%%%%%%%%%%%%%%%%%%%%%%%%%%%%%%%%%%%%%%%%%%%%%
%%%%%%% Talk about FOM - spin Hall angle/effect of Fe%%%%%%%
%%%%%%%%%%%%%%%%%%%%%%%%%%%%%%%%%%%%%%%%%%%%%%%%%%%%%%%%%%%%
The often-used figure of merit for charge-to-spin conversion efficiency is the so-called spin torque efficiency $\xi$. Notably, this metric captures all possible spin current generation mechanisms and is not only limited to the bulk spin Hall effect. The harmonic Hall measurement and the ST-FMR measurement separately and independently provide a measure of $\xi$. From the harmonic Hall measurement, assuming perfect interface transparency, i.e, the magnet absorbs all generated spins or all transferred momentum, the effective spin torque efficiency follows\cite{Avci2014, Roschewsky2017} $\xi_{DL}=2\mid e\mid M_{s}t_{FM}\mu_{0}H_{DL}/(\hbar j_{NM})$, where $M_{s}$ is the saturation magnetization of Co (Extended Data Fig. 3), $t_{FM}$ is the thickness of Co, $\mu_{0} H_{DL}$ is the damping-like torque effective field strength and $j_{NM}$ is the current density flowing through the \FeSi\ layer (Methods). Notably, this expression captures the amount of spin current generated per charge current injected in the \FeSi\ layer. From the analysis, we obtained a $\xi_{DL}$ of 11.92 \% for a-\FefSis\ / Co and 3.98 \% for a-\FefSif\ / Co (Table 1).

From the ST-FMR technique, the $\xi_{DL}$ can be calculated from the expression\cite{Liu2011} $\xi_{DL} = (S/A)(\mid e\mid\mu_{0}M_{s}t_{FM}t_{NM}/\hbar)\sqrt{1+(4\pi M_{eff}/H_{res})}$, where $S/A$ is the ratio between the symmetric ($V_{s}$) and antisymmetric ($V_{A}$) Lorentzian of $V_{mix}$, $\mu_{0}$ is the vacuum permeability, $t_{NM}$ is the thickness of  the a-\FeSi, $H_{res}$ is the resonance field and $4\pi M_{eff}$ is the demagnetization field obtained by fitting to the Kittel relation (Extended Data Fig. 6). This method assumes that the $V_{A}$ is dominantly attributed to the FL-like torque due to Oersted field which under the presence of strong FL-like SOT may obscure the accuracy of the $\xi_{DL}$. Since the FL-like SOT effective field and Oersted field are confirmed to be in the same direction from the harmonic Hall measurement, this method actually serves as a lower bound of $\xi_{DL}$. Historically, it has also been shown that different methods of $\xi_{DL}$ extraction with ST-FMR yielded values less than one order of magnitude difference\cite{Wang2015}. The accurate $\xi_{DL}$ could vary between techniques, nevertheless, the ST-FMR serves the main purpose to confirm the presence of SOT in \FeSi\ / Co. From the analysis, we obtained a $\xi_{DL}$ of 4.30 \% for a-\FefSis\ / Co and 2.56 \% for a-\FefSif\ / Co (Table 1). Finally, to probe the effect of Fe in the a-\FeSi\ layer with respect to the SOT, a control experiment was performed using heavily phosphorous-doped amorphous silicon with no iron doping. No discernible spin-orbit torque signature could be found in either the ST-FMR (Extended Data Fig. 7) or in the harmonic Hall measurements (Extended Data Fig. 8). Therefore, the presence of Fe is clearly necessary for spin current generation in a-\FeSi. 

%%%%%%%%%%%%%%%%%%%%%%%%%%%%%%%%%%%%%%%%%%%%%%%%%%%%%%%%%%%%
%%% Discussion and emphasis on elimination of bulk effect
%%%%%%%%%%%%%%%%%%%%%%%%%%%%%%%%%%%%%%%%%%%%%%%%%%%%%%%%%%%%

Recent work has shown that an intrinsic contribution to spin current is possible from the anomalous Hall effect\cite{Bianco2014, Karel2016}. However, this not only requires a fixed ferromagnet as spin source but also utilizes ferromagnet-spacer-ferromagnet structure\cite{Bose2018, Gibbons2018} and the torque follows the symmetry of a spin accumulation in the magnetization direction of the fixed ferromagnet. Both device configuration and torque symmetry are completely different from our experiments (see Methods for more details). On the other hand, the torque symmetry observed in our experiments are well aligned with what would be expected from a giant SHE and/or Rashba-like interfacial torque. However, the conventional theory describing the intrinsic contribution to the SHE cannot be directly applied as it is formulated based on crystal momentum, \textbf{k}, which is no longer a good quantum number in amorphous materials due to the absence of lattice periodicity.  This point, together with the fact that neither of the elements, Fe or Si, are heavy, make this work very distinct from existing reports on spin orbit torque and spin current generation. Historically, it has been very challenging to cleanly separate out effects owing to the broken symmetry at the interface from those coming from bulk bandstructure of the SOT metal. By contrast, in our experiments any effect coming from the bulk bandstructure is expected to be minimal.

%%%%%%%%%%%%%%%%%%%%%%%%%%%%%%
%%%%%% Conclusion 
%%%%%%%%%%%%%%%%%%%%%%%%%%%%%%
To summarize, we have demonstrated that amorphous \FeSi\ can generate a SOT of similar magnitude as prototypical heavy metals such as Ta\cite{Liu2012} or Pt\cite{Liu2011}. The observed SOT does not conform directly to the conventional theories of spin current generation\cite{Manchon2019}, indicating a potentially new mechanism for spin-orbit torque. This work also extends spin current generation to a completely new class of materials, i.e. silicides, which is ubiquitously used in CMOS technology. The diverse set of available silicides provides an exciting playground for investigating the underlying physics of spin current generation as well as potential for use in integrated Si technology.

%%%%%%%%%%% Figure 1 Schematics and TEM %%%%%%%%%%
\clearpage
\begin{figure}
\begin{centering}
\includegraphics[width=0.9\columnwidth]{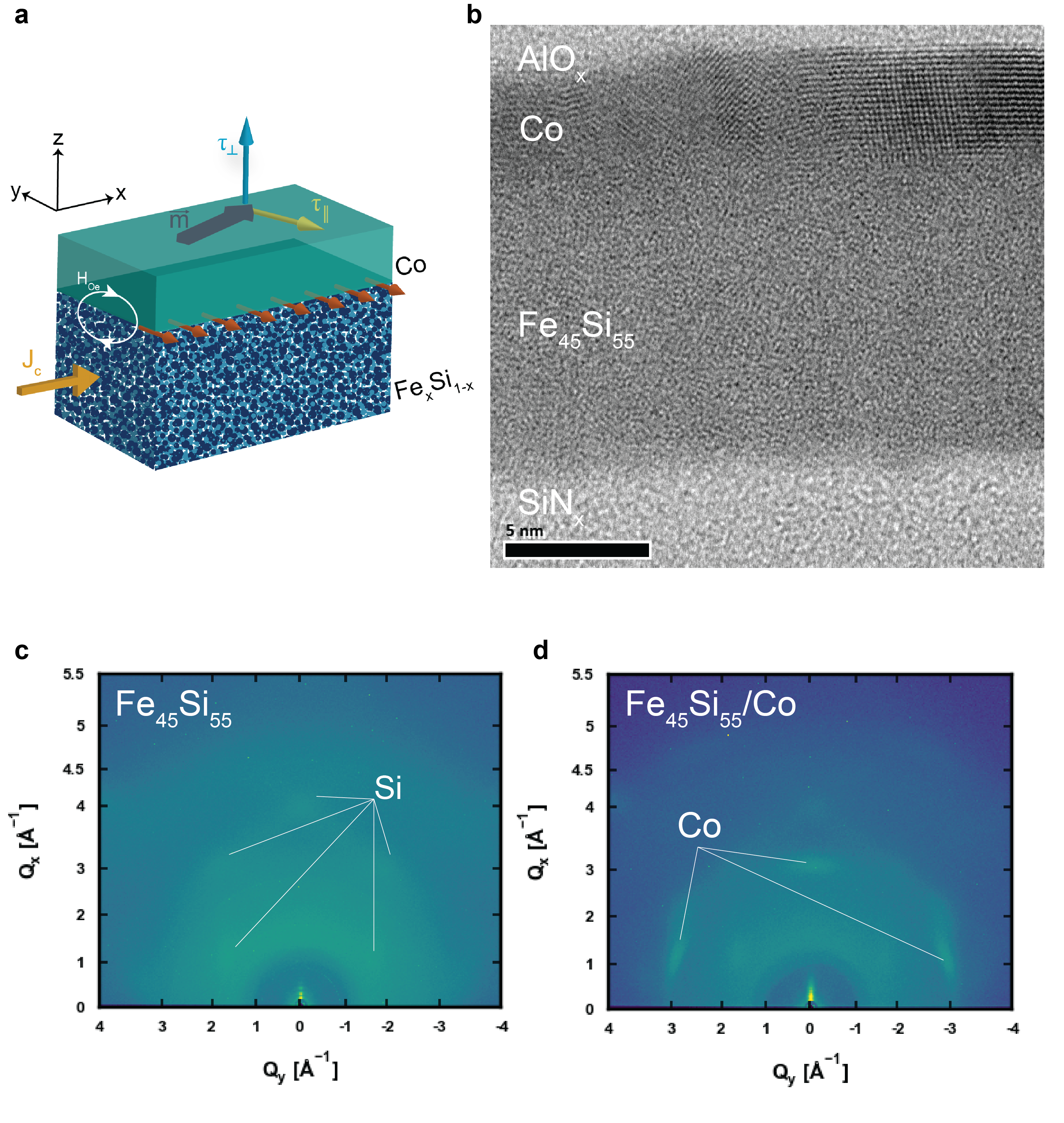}
\end{centering}
\end{figure}

\paragraph*{Fig. 1. Structural characterization of the \FeSi\ / cobalt bilayer with high-resolution transmission electron microscopy and synchrotron X-ray diffraction. }   
{\bf (a)\/} Details of the sample stacks under study. 
{\bf (b)\/} High-resolution cross-sectional TEM of the sample with \FeSi\ (x=45\%). The \FeSi\ layer shows a fully amorphous structure without any nanocrystal formation. The \FeSi\ layer is grown on a 300 nm amorphous silicon nitride on silicon substrate. The cobalt layer is capped with 2 nm oxidized aluminum then 3 nm of silicon nitride (partially shown as the topmost layer).
{\bf (c)\/} Synchrotron X-ray two-dimensional pole figure of bare \FefSif\ (10 nm) on a 300 nm silicon nitride on silicon substrate capped with 2 nm aluminum oxide. Only concentrated intensities corresponding to silicon (1 1 1) and (2 2 0) forming the Kikuchi lines are present. Otherwise, diffuse patterns from the amorphous \FeSi\ show the lack of long-range order.
{\bf (d)\/} Synchrotron X-ray two-dimensional pole figure of \FefSif\ (8 nm) / Co (4 nm) on 300 nm silicon nitride on silicon substrate capped with 2 nm aluminum oxide. On top of the silicon (1 1 1) and (2 2 0) peaks, additional sharp cobalt peaks are the only contrast to the \FefSif\ bare film 2D pole figure.

%%%%%%%%%%% Figure 2 harmonic Hall %%%%%%%%%%%%
\clearpage
\begin{figure}
\begin{centering}
\includegraphics[width=0.9\columnwidth]{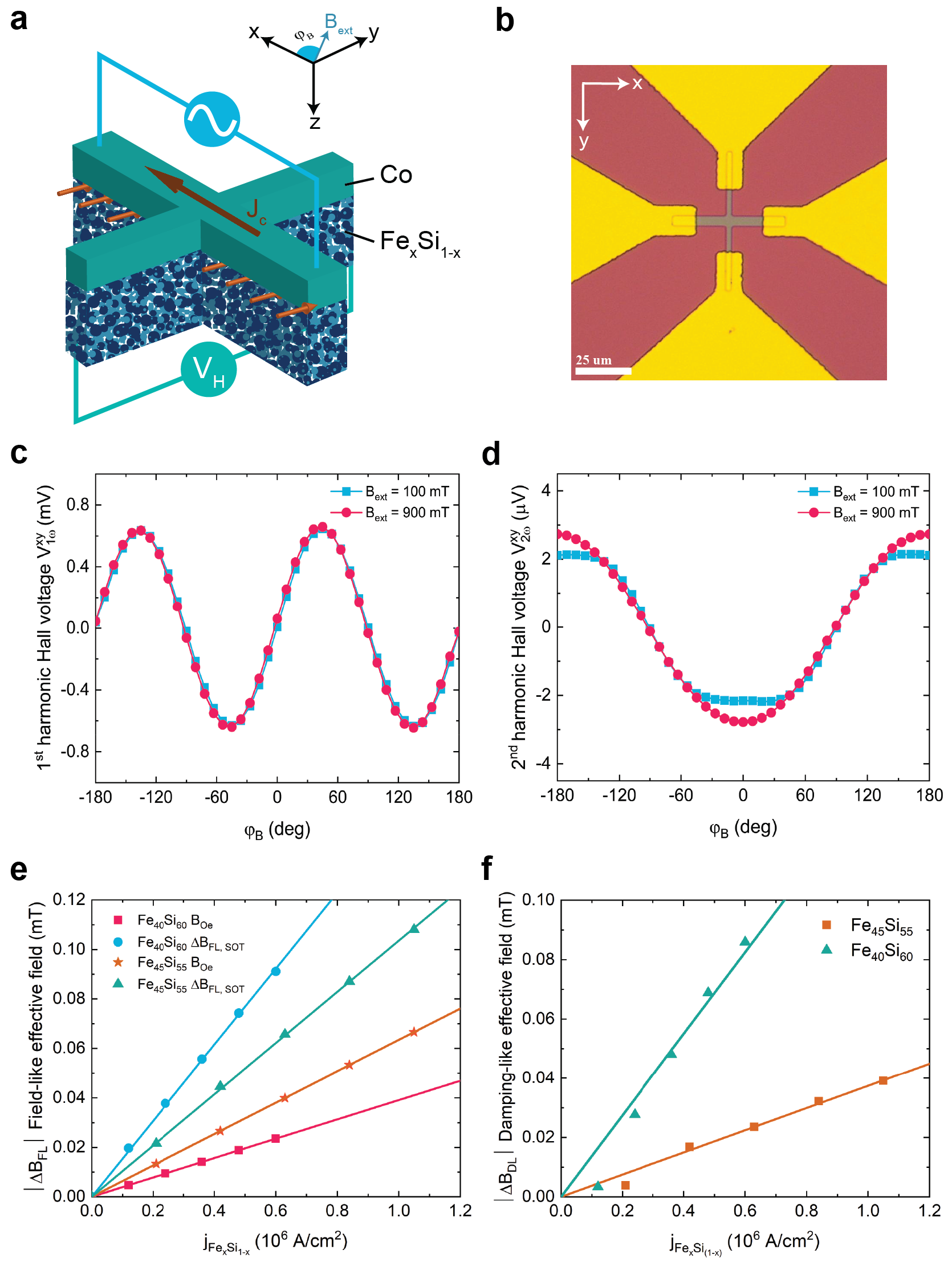}
\end{centering}
\end{figure}

\paragraph*{Fig. 2. Harmonic Hall measurement of \FeSi\ / cobalt bilayer.}   
{\bf (a)\/} Schematic of the harmonic Hall measurement setup. Low frequency a.c. current is applied along the x-direction and the magnetic moments of cobalt are pinned by a rotating external magnetic field in the xy plane.
{\bf (b)\/} Optical microscope image of the fabricated single Hall bar device. The current line is 5 micron in width and the voltage line is 3 micron in width. Both lines are 30 micron in length. 
{\bf (c)\/} First harmonic voltage of the \FefSif\ / cobalt bilayer. The first harmonic voltage is essentially the planar Hall voltage that scales as sin(2$\varphi_{B}$). 
{\bf (d)\/} Second harmonic voltage of the \FefSif\ / cobalt bilayer. The second harmonic voltage contains information on field-like (FL) torque effective field, damping-like (DL) torque effective field and thermoelectric effects. The FL-like torque effective field scales as cos(2$\varphi_{B}$)cos($\varphi_{B}$). The DL-like torque effective field and thermoelectric effect scale as cos($\varphi_{B}$). 
{\bf (e)\/} FL-like torque effective field due to SOT and Oersted field as a function of injected current density in the \FeSi\ layer. The linear trend confirms the presence of appreciable FL-like spin-orbit torque.
{\bf (f)\/} DL-like torque effective field as a function of injected current density in the \FeSi\ layer. The linear trend confirms the presence of appreciable DL-like spin-orbit torque.

%%%%%%%%%%% Figure 3  ST-FMR %%%%%%%%%%%
\clearpage
\begin{figure}
\begin{centering}
\includegraphics[width=0.92\columnwidth]{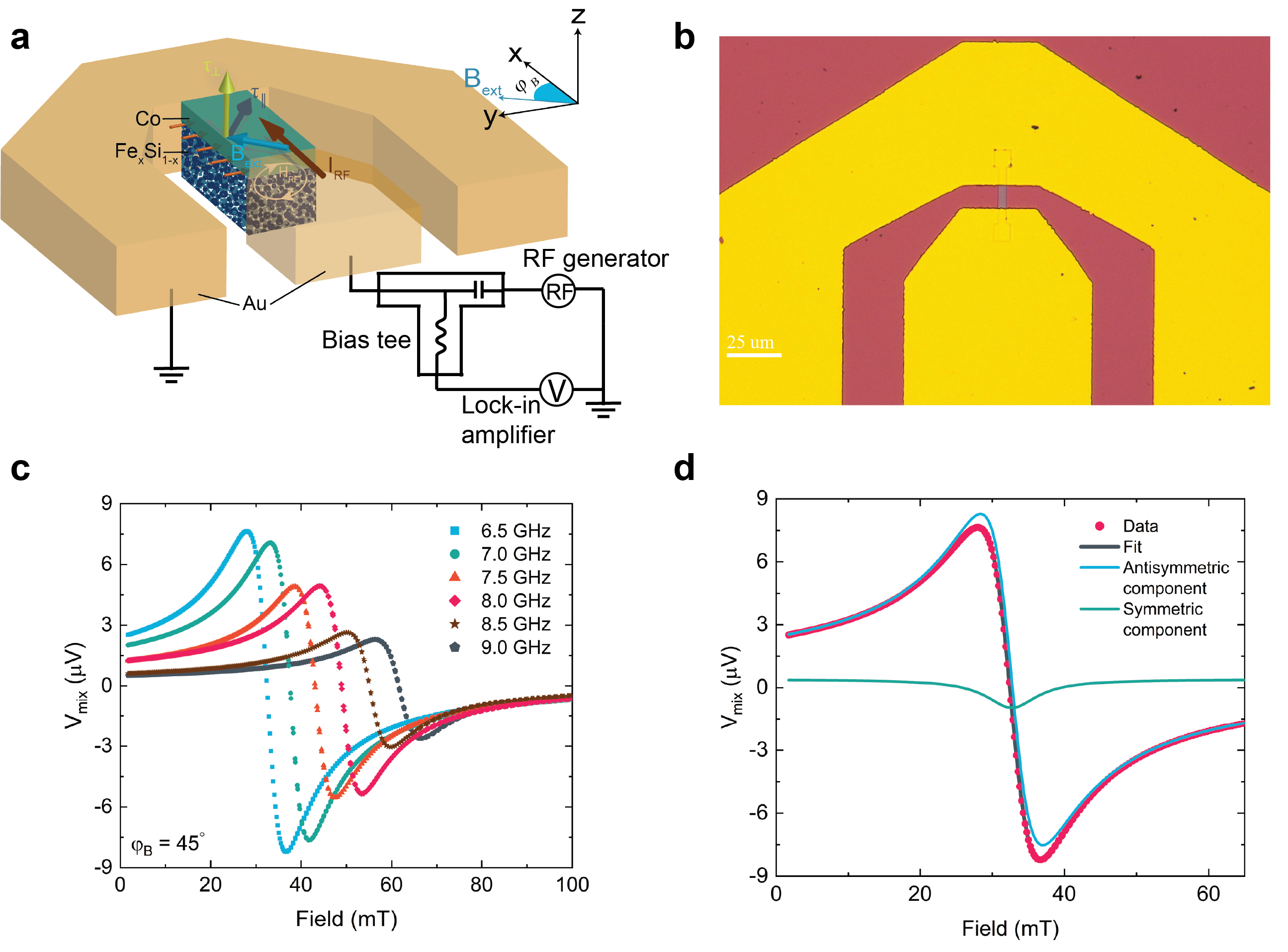}
\end{centering}
\end{figure}

\paragraph*{Fig. 3. Spin-torque ferromagnetic resonance (ST-FMR) measurement of \FefSif\ (8) / cobalt (4) bilayer.}   
{\bf (a)\/} Schematic of the ST-FMR setup and sample geometry under study. $\tau_{\parallel}$ is the in-plane current-induced torque component, $\tau_{\perp}$ is the out-of-plane current-induced torque component and $\varphi_{B}$ is the angle spanned between the in-plane field and RF current direction.
{\bf (b)\/} Optical microscope image of fabricated ST-FMR device with dimension of 3 micron in width and 9 micron in length. The coplanar waveguide (gold color) serves as the microwave transmission contact to the device. 
{\bf (c)\/} ST-FMR spectrum measured on \FefSif\ (8) / cobalt (4) versus applied external magnetic field with RF frequency from 6.5 GHz to 9 GHz. 
{\bf (d)\/} ST-FMR $V_{mix}$ lineshape fitting on \FefSif\ (8) / cobalt (4) versus external magnetic field at RF frequency 6.5 GHz (Methods). 

%%%%%%%%%%% Table 1: spin torque efficiency %%%%%%%%%%%
%attached as figure
\clearpage
\begin{figure}
\begin{centering}
\includegraphics[width=0.9\columnwidth]{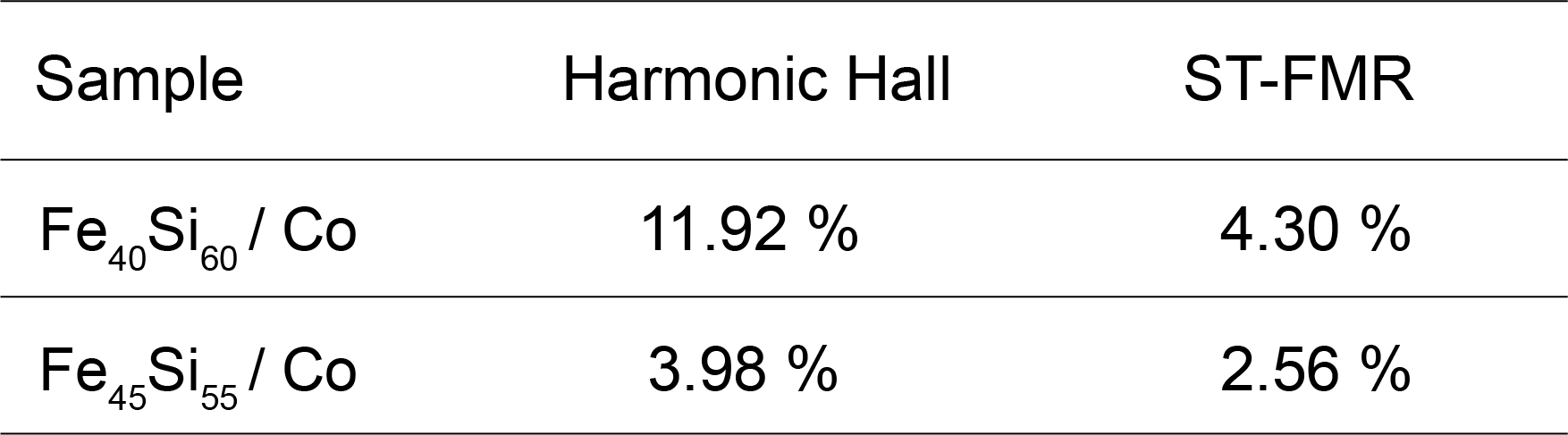}
\end{centering}
\end{figure}
\paragraph*{Table 1. Spin torque efficiency $\mid\xi_{DL}\mid$ extracted from harmonic Hall measurement and ST-FMR on \FeSi\ / cobalt bilayer.}

%%%%%%%%%%%%%%%%%%%%%%%%%%%%%%
%%%% References 
%%%%%%%%%%%%%%%%%%%%%%%%%%%%%%
\clearpage
\bibliographystyle{naturemag}
\bibliography{reference}

\begin{thebibliography}{10}
\expandafter\ifx\csname url\endcsname\relax
  \def\url#1{\texttt{#1}}\fi
\expandafter\ifx\csname urlprefix\endcsname\relax\def\urlprefix{URL }\fi
\providecommand{\bibinfo}[2]{#2}
\providecommand{\eprint}[2][]{\url{#2}}

\bibitem{Liu2011}
\bibinfo{author}{Liu, L.}, \bibinfo{author}{Moriyama, T.},
  \bibinfo{author}{Ralph, D.~C.} \& \bibinfo{author}{Buhrman, R.~A.}
\newblock \bibinfo{title}{Spin-torque ferromagnetic resonance induced by the
  spin hall effect}.
\newblock \emph{\bibinfo{journal}{Physical Review Letters}}
  \textbf{\bibinfo{volume}{106}}, \bibinfo{pages}{036601}
  (\bibinfo{year}{2011}).

\bibitem{Miron2011}
\bibinfo{author}{Miron, I.~M.} \emph{et~al.}
\newblock \bibinfo{title}{Perpendicular switching of a single ferromagnetic
  layer induced by in-plane current injection}.
\newblock \emph{\bibinfo{journal}{Nature}} \textbf{\bibinfo{volume}{476}},
  \bibinfo{pages}{189--193} (\bibinfo{year}{2011}).

\bibitem{Liu2012}
\bibinfo{author}{Liu, L.} \emph{et~al.}
\newblock \bibinfo{title}{Spin-torque switching with the giant spin hall effect
  of tantalum}.
\newblock \emph{\bibinfo{journal}{Science}} \textbf{\bibinfo{volume}{336}},
  \bibinfo{pages}{555--558} (\bibinfo{year}{2012}).

\bibitem{Mellnik2014}
\bibinfo{author}{Mellnik, A.~R.} \emph{et~al.}
\newblock \bibinfo{title}{Spin-transfer torque generated by a topological
  insulator}.
\newblock \emph{\bibinfo{journal}{Nature}} \textbf{\bibinfo{volume}{511}},
  \bibinfo{pages}{449--451} (\bibinfo{year}{2014}).

\bibitem{Hayashi2014}
\bibinfo{author}{Hayashi, M.}, \bibinfo{author}{Kim, J.},
  \bibinfo{author}{Yamanouchi, M.} \& \bibinfo{author}{Ohno, H.}
\newblock \bibinfo{title}{Quantitative characterization of the spin-orbit
  torque using harmonic hall voltage measurements}.
\newblock \emph{\bibinfo{journal}{Physical Review B}}
  \textbf{\bibinfo{volume}{89}}, \bibinfo{pages}{144425}
  (\bibinfo{year}{2014}).

\bibitem{Avci2014}
\bibinfo{author}{Avci, C.~O.} \emph{et~al.}
\newblock \bibinfo{title}{Interplay of spin-orbit torque and thermoelectric
  effects in ferromagnet/normal-metal bilayers}.
\newblock \emph{\bibinfo{journal}{Physical Review B}}
  \textbf{\bibinfo{volume}{90}}, \bibinfo{pages}{224427}
  (\bibinfo{year}{2014}).

\bibitem{Wang2015}
\bibinfo{author}{Wang, Y.} \emph{et~al.}
\newblock \bibinfo{title}{Topological surface states originated spin-orbit
  torques in bi2se3}.
\newblock \emph{\bibinfo{journal}{Physical Review Letters}}
  \textbf{\bibinfo{volume}{114}}, \bibinfo{pages}{257202}
  (\bibinfo{year}{2015}).

\bibitem{Chen2016}
\bibinfo{author}{Chen, L.} \emph{et~al.}
\newblock \bibinfo{title}{Robust spin-orbit torque and spin-galvanic effect at
  the fe/gaas (001) interface at room temperature}.
\newblock \emph{\bibinfo{journal}{Nature Communications}}
  \textbf{\bibinfo{volume}{7}}, \bibinfo{pages}{13802} (\bibinfo{year}{2016}).

\bibitem{Skinner2015}
\bibinfo{author}{Skinner, T.~D.} \emph{et~al.}
\newblock \bibinfo{title}{Complementary spin-hall and inverse spin-galvanic
  effect torques in a ferromagnet/semiconductor bilayer}.
\newblock \emph{\bibinfo{journal}{Nature Communications}}
  \textbf{\bibinfo{volume}{6}}, \bibinfo{pages}{6730} (\bibinfo{year}{2015}).

\bibitem{Shao2016}
\bibinfo{author}{Shao, Q.} \emph{et~al.}
\newblock \bibinfo{title}{Strong rashba-edelstein effect-induced spin-orbit
  torques in monolayer transition metal dichalcogenide/ferromagnet bilayers}.
\newblock \emph{\bibinfo{journal}{Nano Letters}} \textbf{\bibinfo{volume}{16}},
  \bibinfo{pages}{7514--7520} (\bibinfo{year}{2016}).

\bibitem{MacNeill2017}
\bibinfo{author}{MacNeill, D.} \emph{et~al.}
\newblock \bibinfo{title}{Control of spin-orbit torques through crystal
  symmetry in wte 2 /ferromagnet bilayers}.
\newblock \emph{\bibinfo{journal}{Nature Physics}}
  \textbf{\bibinfo{volume}{13}}, \bibinfo{pages}{300--305}
  (\bibinfo{year}{2017}).

\bibitem{Nan2019}
\bibinfo{author}{Nan, T.} \emph{et~al.}
\newblock \bibinfo{title}{Anisotropic spin-orbit torque generation in epitaxial
  sriro 3 by symmetry design}.
\newblock \emph{\bibinfo{journal}{Proceedings of the National Academy of
  Sciences}} \textbf{\bibinfo{volume}{116}}, \bibinfo{pages}{16186--16191}
  (\bibinfo{year}{2019}).

\bibitem{Manchon2019}
\bibinfo{author}{Manchon, A.} \emph{et~al.}
\newblock \bibinfo{title}{Current-induced spin-orbit torques in ferromagnetic
  and antiferromagnetic systems}.
\newblock \emph{\bibinfo{journal}{Reviews of Modern Physics}}
  \textbf{\bibinfo{volume}{91}}, \bibinfo{pages}{035004}
  (\bibinfo{year}{2019}).

\bibitem{Luo2020}
\bibinfo{author}{Luo, Z.} \emph{et~al.}
\newblock \bibinfo{title}{{Current-driven magnetic domain-wall logic}}.
\newblock \emph{\bibinfo{journal}{Nature}} \textbf{\bibinfo{volume}{579}},
  \bibinfo{pages}{214--218} (\bibinfo{year}{2020}).

\bibitem{Noel2020}
\bibinfo{author}{No{\"{e}}l, P.} \emph{et~al.}
\newblock \bibinfo{title}{{Non-volatile electric control of spin–charge
  conversion in a SrTiO3 Rashba system}}.
\newblock \emph{\bibinfo{journal}{Nature}} \textbf{\bibinfo{volume}{580}},
  \bibinfo{pages}{483--486} (\bibinfo{year}{2020}).

\bibitem{Tsai2020}
\bibinfo{author}{Tsai, H.} \emph{et~al.}
\newblock \bibinfo{title}{{Electrical manipulation of a topological
  antiferromagnetic state}}.
\newblock \emph{\bibinfo{journal}{Nature}} \textbf{\bibinfo{volume}{580}},
  \bibinfo{pages}{608--613} (\bibinfo{year}{2020}).

\bibitem{Ando2012}
\bibinfo{author}{Ando, K.} \& \bibinfo{author}{Saitoh, E.}
\newblock \bibinfo{title}{{Observation of the inverse spin Hall effect in
  silicon}}.
\newblock \emph{\bibinfo{journal}{Nature Communications}}
  \textbf{\bibinfo{volume}{3}}, \bibinfo{pages}{629} (\bibinfo{year}{2012}).

\bibitem{Roschewsky2017}
\bibinfo{author}{Roschewsky, N.}, \bibinfo{author}{Lambert, C.~H.} \&
  \bibinfo{author}{Salahuddin, S.}
\newblock \bibinfo{title}{{Spin-orbit torque switching of ultralarge-thickness
  ferrimagnetic GdFeCo}}.
\newblock \emph{\bibinfo{journal}{Physical Review B}}
  \textbf{\bibinfo{volume}{96}}, \bibinfo{pages}{1--5} (\bibinfo{year}{2017}).

\bibitem{Roschewsky2019}
\bibinfo{author}{Roschewsky, N.} \emph{et~al.}
\newblock \bibinfo{title}{Spin-orbit torque and nernst effect in bi-sb/co
  heterostructures}.
\newblock \emph{\bibinfo{journal}{Physical Review B}}
  \textbf{\bibinfo{volume}{99}}, \bibinfo{pages}{195103--195104}
  (\bibinfo{year}{2019}).

\bibitem{Hirsch1999}
\bibinfo{author}{Hirsch, J.~E.}
\newblock \bibinfo{title}{Spin hall effect}.
\newblock \emph{\bibinfo{journal}{Phys. Rev. Lett.}}
  \textbf{\bibinfo{volume}{83}}, \bibinfo{pages}{1834} (\bibinfo{year}{1999}).

\bibitem{Sinova2015}
\bibinfo{author}{Sinova, J.}, \bibinfo{author}{Valenzuela, S.~O.},
  \bibinfo{author}{Wunderlich, J.}, \bibinfo{author}{Back, C.~H.} \&
  \bibinfo{author}{Jungwirth, T.}
\newblock \bibinfo{title}{Spin hall effects}.
\newblock \emph{\bibinfo{journal}{Reviews of Modern Physics}}
  \textbf{\bibinfo{volume}{87}}, \bibinfo{pages}{1213--1260}
  (\bibinfo{year}{2015}).

\bibitem{Zhang2003}
\bibinfo{author}{Zhang, S.-L.} \& \bibinfo{author}{{\"{O}}stling, M.}
\newblock \bibinfo{title}{{Metal Silicides in CMOS Technology: Past, Present,
  and Future Trends}}.
\newblock \emph{\bibinfo{journal}{Critical Reviews in Solid State and Materials
  Sciences}} \textbf{\bibinfo{volume}{28}}, \bibinfo{pages}{1--129}
  (\bibinfo{year}{2003}).

\bibitem{Chen2005}
\bibinfo{author}{Chen, L.~J.}
\newblock \bibinfo{title}{{Metal silicides: An integral part of
  microelectronics}}.
\newblock \emph{\bibinfo{journal}{JOM}} \textbf{\bibinfo{volume}{57}},
  \bibinfo{pages}{24--30} (\bibinfo{year}{2005}).

\bibitem{Ren2011}
\bibinfo{author}{Ren, L.} \& \bibinfo{author}{Tu, K.}
\newblock \bibinfo{title}{{Silicide technology for SOI devices}}.
\newblock In \emph{\bibinfo{booktitle}{Silicide Technology for Integrated
  Circuits}}, chap.~\bibinfo{chapter}{8}, \bibinfo{pages}{201--228}
  (\bibinfo{publisher}{IET}, \bibinfo{address}{The Institution of Engineering
  and Technology, Michael Faraday House, Six Hills Way, Stevenage SG1 2AY, UK},
  \bibinfo{year}{2011}).

\bibitem{Salahuddin2018}
\bibinfo{author}{Salahuddin, S.}, \bibinfo{author}{Ni, K.} \&
  \bibinfo{author}{Datta, S.}
\newblock \bibinfo{title}{The era of hyper-scaling in electronics}.
\newblock \emph{\bibinfo{journal}{Nature Electronics}}
  \textbf{\bibinfo{volume}{1}}, \bibinfo{pages}{442--450}
  (\bibinfo{year}{2018}).

\bibitem{Dyakonov1971}
\bibinfo{author}{Dyakonov, M.~I.} \& \bibinfo{author}{Perel, V.~I.}
\newblock \bibinfo{title}{{Current-induced spin orientation of electrons in
  semiconductors}}.
\newblock \emph{\bibinfo{journal}{Physics Letters A}}
  \textbf{\bibinfo{volume}{35}}, \bibinfo{pages}{459--460}
  (\bibinfo{year}{1971}).

\bibitem{Edelstein1990}
\bibinfo{author}{Edelstein, V.~M.}
\newblock \bibinfo{title}{Spin polarization of conduction electrons induced by
  electric current in two-dimensional asymmetric electron systems}.
\newblock \emph{\bibinfo{journal}{Solid State Communications}}
  \textbf{\bibinfo{volume}{73}}, \bibinfo{pages}{233--235}
  (\bibinfo{year}{1990}).

\bibitem{Hellman2017}
\bibinfo{author}{Hellman, F.} \emph{et~al.}
\newblock \bibinfo{title}{Interface-induced phenomena in magnetism}.
\newblock \emph{\bibinfo{journal}{Reviews of Modern Physics}}
  \textbf{\bibinfo{volume}{89}}, \bibinfo{pages}{025006}
  (\bibinfo{year}{2017}).

\bibitem{Slonczewski1996}
\bibinfo{author}{Slonczewski, J.~C.}
\newblock \bibinfo{title}{Current-driven excitation of magnetic multilayers}.
\newblock \emph{\bibinfo{journal}{Journal of Magnetism and Magnetic Materials}}
  \textbf{\bibinfo{volume}{159}}, \bibinfo{pages}{L1--L7}
  (\bibinfo{year}{1996}).

\bibitem{Ralph2008}
\bibinfo{author}{Ralph, D.~C.} \& \bibinfo{author}{Stiles, M.~D.}
\newblock \bibinfo{title}{Spin transfer torques}.
\newblock \emph{\bibinfo{journal}{Journal of Magnetism and Magnetic Materials}}
  \textbf{\bibinfo{volume}{320}}, \bibinfo{pages}{1190--1216}
  (\bibinfo{year}{2008}).

\bibitem{Manchon2008}
\bibinfo{author}{Manchon, A.} \& \bibinfo{author}{Zhang, S.}
\newblock \bibinfo{title}{Theory of nonequilibrium intrinsic spin torque in a
  single nanomagnet}.
\newblock \emph{\bibinfo{journal}{Physical Review B}}
  \textbf{\bibinfo{volume}{78}} (\bibinfo{year}{2008}).

\bibitem{Sharma1975}
\bibinfo{author}{Sharma, S.~K.}, \bibinfo{author}{Geserich, H.~P.} \&
  \bibinfo{author}{Theiner, W.~A.}
\newblock \bibinfo{title}{{Amorphous‐to‐Crystalline Transformation in
  Evaporated FeSi Films}}.
\newblock \emph{\bibinfo{journal}{physica status solidi (a)}}
  \textbf{\bibinfo{volume}{32}}, \bibinfo{pages}{467--474}
  (\bibinfo{year}{1975}).

\bibitem{Karel2018}
\bibinfo{author}{Karel, J.} \emph{et~al.}
\newblock \bibinfo{title}{{Enhanced spin polarization of amorphous F ex S i1-x
  thin films revealed by Andreev reflection spectroscopy}}.
\newblock \emph{\bibinfo{journal}{Physical Review Materials}}
  \textbf{\bibinfo{volume}{2}}, \bibinfo{pages}{1--6} (\bibinfo{year}{2018}).

\bibitem{Bianco2014}
\bibinfo{author}{Bianco, R.}, \bibinfo{author}{Resta, R.} \&
  \bibinfo{author}{Souza, I.}
\newblock \bibinfo{title}{How disorder affects the berry-phase anomalous hall
  conductivity: A reciprocal-space analysis}.
\newblock \emph{\bibinfo{journal}{Physical Review B}}
  \textbf{\bibinfo{volume}{90}}, \bibinfo{pages}{125153}
  (\bibinfo{year}{2014}).

\bibitem{Karel2016}
\bibinfo{author}{Karel, J.} \emph{et~al.}
\newblock \bibinfo{title}{{Scaling of the anomalous Hall effect in lower
  conductivity regimes}}.
\newblock \emph{\bibinfo{journal}{EPL (Europhysics Letters)}}
  \textbf{\bibinfo{volume}{114}}, \bibinfo{pages}{57004}
  (\bibinfo{year}{2016}).

\bibitem{Bose2018}
\bibinfo{author}{Bose, A.} \emph{et~al.}
\newblock \bibinfo{title}{{Observation of Anomalous Spin Torque Generated by a
  Ferromagnet}}.
\newblock \emph{\bibinfo{journal}{Physical Review Applied}}
  \textbf{\bibinfo{volume}{9}}, \bibinfo{pages}{064026} (\bibinfo{year}{2018}).

\bibitem{Gibbons2018}
\bibinfo{author}{Gibbons, J.~D.}, \bibinfo{author}{MacNeill, D.},
  \bibinfo{author}{Buhrman, R.~A.} \& \bibinfo{author}{Ralph, D.~C.}
\newblock \bibinfo{title}{{Reorientable Spin Direction for Spin Current
  Produced by the Anomalous Hall Effect}}.
\newblock \emph{\bibinfo{journal}{Physical Review Applied}}
  \textbf{\bibinfo{volume}{9}}, \bibinfo{pages}{064033} (\bibinfo{year}{2018}).

\end{thebibliography}
\clearpage

%%%%%%%%%%%%%%%%%%%%%%%%%%%%%%%%%%%%%%%%%%%%%%%%%%%%%%%%%%%%%%%%
%%%% END MATTER (Ackn.+ Corspdnce)
%%%%%%%%%%%%%%%%%%%%%%%%%%%%%%%%%%%%%%%%%%%%%%%%%%%%%%%%%%%%%%%%
\newpage
\begin{addendum}

\item [Acknowledgements] 
This work was primarily supported by the U.S. Department of Energy, Office of Science, Office of Basic Energy Sciences, Materials Sciences and Engineering Division, under Contract No. DE-AC02-05-CH11231 within the Non-Equilibrium Magnetism program (MSMAG). This work was also supported in part by the ASCENT center, one of the six centers within the JUMP initiative jointly supported by DARPA and SRC. In addition, support from NSF E3S center is gratefully acknowledged. This work was also supported through the Australian Research Council Centre of Excellence in Future Low Energy Electronics Technologies under CE170100039 and the Australian Research Council Discovery Project DP200102477. We acknowledge access to NCRIS facilities (ANFF and the Heavy Ion Accelerator Capability) at the Australian National University. Use of the Stanford Synchrotron Radiation Lightsource, SLAC National Accelerator Laboratory, is supported by the U.S. Department of Energy, Office of Science, Office of Basic Energy Sciences under Contract No. DE-AC02-76SF00515. This work was performed in part at the Berkeley Marvell Nanofabrication Laboratory at University of California, Berkeley and their staff research support is greatly appreciated.

\item [Correspondence] 
Correspondence and requests for materials should be addressed to C.-H.H. (chhsu@berkeley.edu) or S.S. (sayeef@berkeley.edu).

\end{addendum}

\end{document}